\documentclass[%
reprint,
amsmath,amssymb,
aps,
]{revtex4-1}

\usepackage{graphicx}
\usepackage{dcolumn}
\usepackage{bm}
\usepackage{hyperref}
\usepackage{siunitx}
\usepackage{amsmath}
\usepackage{mathtools}

\begin{document}


\title{Towards nanoelectromechanical photothermal localization microscopy}

\author{Miao-Hsuan Chien}
\affiliation{Institute of Sensor and Actuator Systems, TU Wien, Gusshausstrasse 27-29, 1040 Vienna, Austria.}

\author{Silvan Schmid}
\email{silvan.schmid@tuwien.ac.at}
\affiliation{Institute of Sensor and Actuator Systems, TU Wien, Gusshausstrasse 27-29, 1040 Vienna, Austria.}

\date{\today}

\begin{abstract}
Single-molecule microscopy has become an indispensable tool for biochemical analysis. The capability of characterizing distinct properties of individual molecules without averaging has provided us with a different perspective for the existing scientific issues and phenomena. Recently, the super-resolution fluorescence microscopy techniques have overcome the optical diffraction limit by the localization of molecule positions. However, the labelling process can potentially modify the intermolecular dynamics. Based on the highly-sensitive nanomechanical photothermal microscopy reported previously, we propose optimizations on this label-free microscopy technique towards localization microscopy. A localization precision of \SI{3}{\angstrom} is achieved with gold nanoparticles, and the detection of polarization-dependent absorption is demonstrated, which opens the door for further improvement with polarization modulation imaging.
\end{abstract}


\maketitle

\section{Introduction}

Single-molecule microscopy has enabled the precise detection of individual characteristics without averaging for biochemical traces. Label-free single-molecule imaging further offers the possibility of detecting the authentic system dynamics without the modifications of the intermolecular interactions resulting from labeling \cite{liang2017photonic}. Label-free techniques also bypass the photobleaching issue of the fluorescent dyes. To achieve a high sensitivity with label-free optical microscopy, detecting the absorption of nano-objects has the advantage over detecting the scattering, since the absorption cross-sections scale linearly with its volume while the scattering cross-sections scale quadratically \cite{bohren2008absorption}.  Absorption-based optical microscopy with single-molecule sensitivity has been demonstrated by means of transmission microscopy \cite{kukura2010single,celebrano2011single},  ground-state depletion microscopy \cite{chong2010ground} and photothermal microscopy \cite{gaiduk2010detection,gaiduk2010room,berciaud2004photothermal,cognet2008photothermal,nedosekin2014super,chang2012enhancing,ding2016hundreds}.  Among these techniques, photothermal microscopy measures the direct absorption of molecules from their photothermal heating instead of the relative attenuation of incident light in the ppm regime, and thus can provide even higher signal-to-noise ratio. With the recent discoveries on responsive imaging media, the sensitivity of photothermal contrast microscopy has been pushed to \SI{}{\pico\watt\per\sqrt{\hertz}} regime with near-critical Xenon \cite{ding2016hundreds}. 

\begin{figure}
	\centering
	\includegraphics[width=0.3\textwidth]{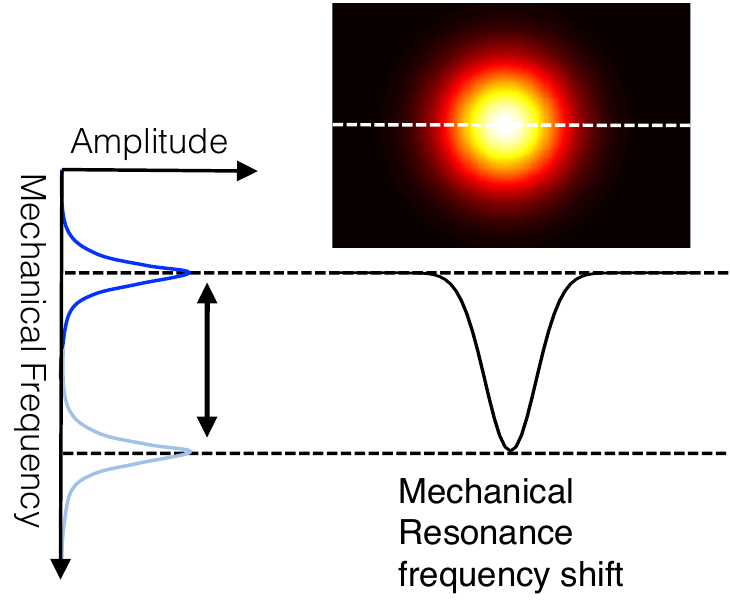} 
	\caption{The working principle of NEMS photothermal microscopy.}
	\label{fig:fig0} 
\end{figure}

As an alternative for the conventional photothermal microscopy that relies on the temperature-sensitive refractive index of the medium for imaging, photothermal microscopy using nanomechanical resonators as a temperature-sensitive element has demonstrated unprecedented sensitivity of \SI{16}{\femto\watt\per\sqrt{\hertz}} recently \cite{chien2018single}. An averaged localization precision of \SI{32}{\nano\meter} was extracted from the single-molecule signal. However, in comparison with the localization precision better than \SI{10}{\nano\meter} routinely achieved by the single-molecule localization microscopy techniques \cite{rust2006sub,shroff2008live,rittweger2009sted,mortensen2010optimized}, it's clear that there's still space for improvements. We hereby present an extensive research on the optimizations of nanoelectromechanical (NEMS) photothermal microscopy to achieve better beam profile and localization precision.

\begin{figure*}
	\centering
	\includegraphics[width=0.8\textwidth]{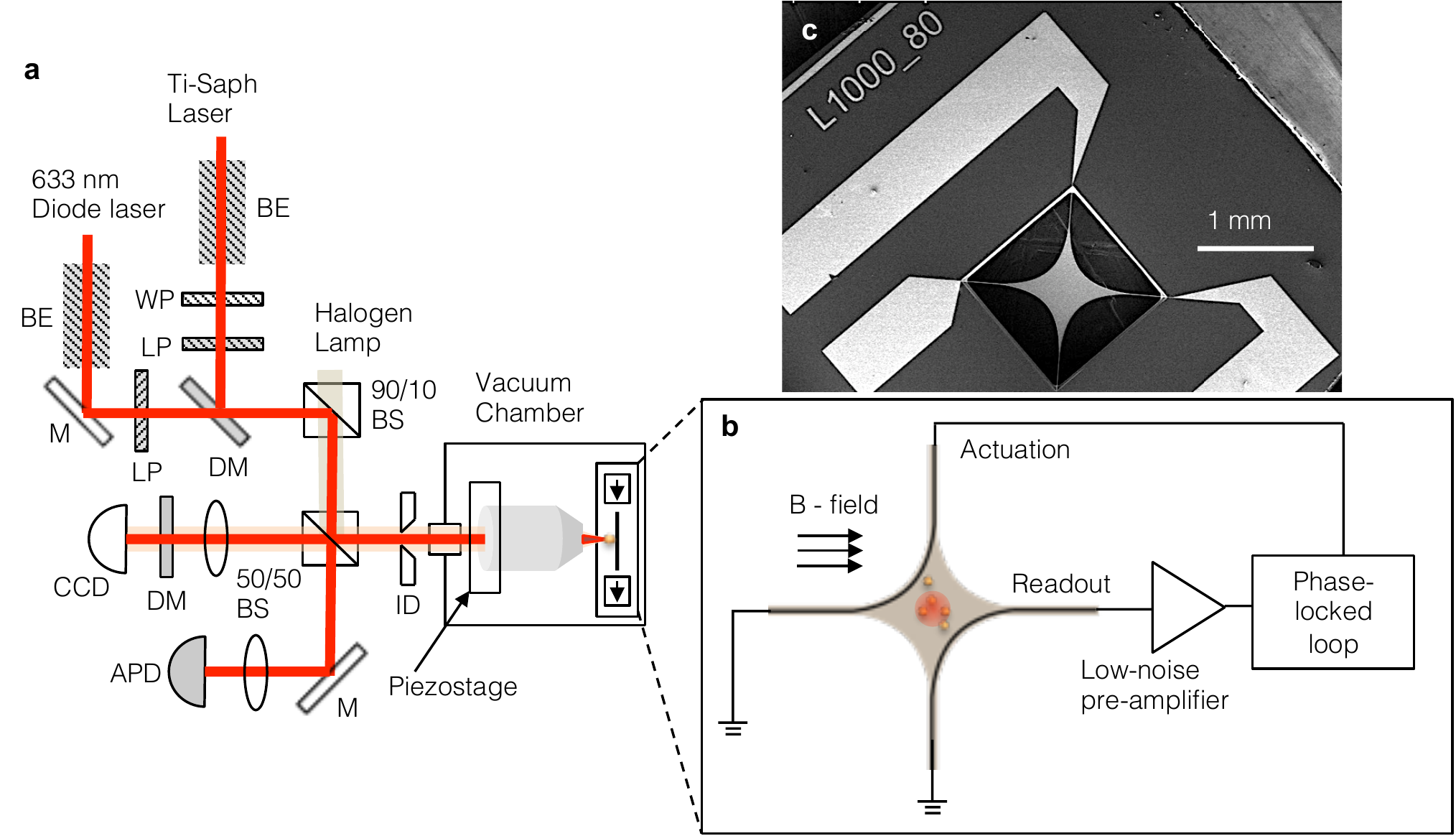} 
	\caption{(a) Schematic of the measurement setup. BE: beam expander. M: mirror. WP: waveplate. LP: linear polarizer. BS: beam splitter. PD: photodetector/powermeter. DM: dichroic mirror. ID: iris diaphram. (b) The transduction scheme of the trampoline resonator. (c) SEM image of the trampoline resonator.}
	\label{fig:fig1} 
\end{figure*}
The working principle of NEMS photothermal microscopy is briefly depicted in Figure~\ref{fig:fig0}. Upon scanning, the photothermal heating of the sample results in a detectable detuning in the mechanical resonance frequency. The image contrast can thus be obtained by tracking the frequency shift of the nanomechanical resonator. Here we report on the following optimizations:
\begin{itemize}
	\item A dedicated optical setup that allows full control of beam diameter, power, polarization, and alignment is established to replace the fixed laser source of the previously used laser-Doppler vibrometer (LDV). We this we achieve an optimal beam quality, as shown in the setup schematic in Figure~\ref{fig:fig1}a. A diode laser with \SI{633}{\nano\meter} wavelength (Toptica TopMode) and a Titanium-Sapphire laser (M Squared SolsTiS) with wavelength locked at \SI{800}{\nano\meter} are used. Both lasers are linearly polarized. 
	\item Instead of the optical vibrometric readout with a commercial LDV, an integrated inductive transduction scheme for both readout and actuation is implemented to provide more flexibility and compatibility with the optical setup, as shown in Figure~\ref{fig:fig1}b. The movement of the gold electrode in the static magnetic field results in an alternating voltage that is first amplified with a low-noise pre-amplifier and fed to the lock-in amplifier, and the frequency is tracked with the phase-locked loop (PLL) (HF2LI, Zurich Instrument). An enhanced Halbach array was used to create magnetic field of around \SI{1}{\tesla} over the center distance of \SI{5}{\milli\meter}. 
	\item The scanning is done with a closed-loop piezoelectric nanopositioning stage (PiMars, Physikinstrumente) with \SI{2}{\nano\meter} resolution to provide finer imaging data and thus better localization.
	\item Instead of nanomechanical drums used in the previous work, NEMS trampoline resonators with a large center area for scanning are used in the present work, as shown in Figure~\ref{fig:fig1}c. Trampolines maintain sufficient area for imaging while providing higher responsivity with the same initial stress and window size, benefiting from the good thermal isolation of the thin tethers.
\end{itemize}

\section{Results and Discussion}
\begin{figure*}
	\centering
	\includegraphics[width=0.9\textwidth]{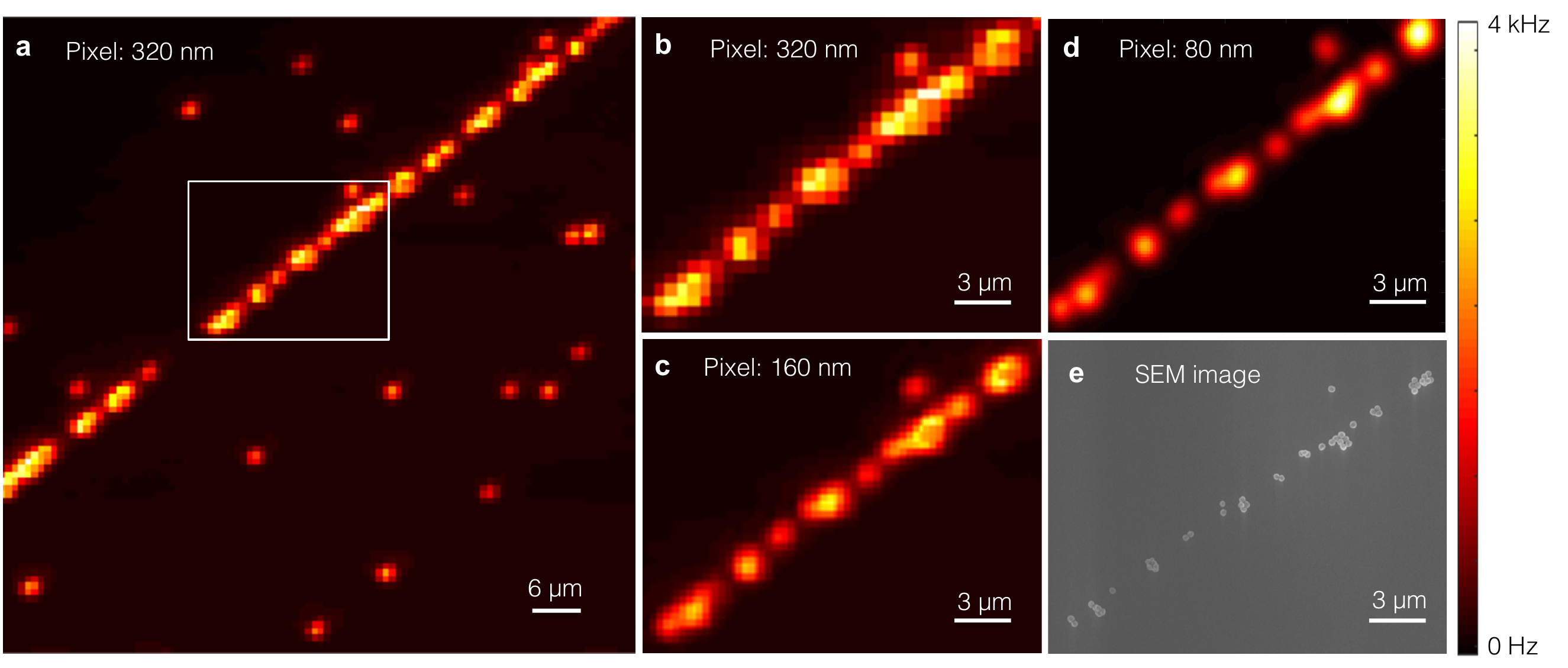} 
	\caption{(a) NEMS photothermal microscopy image with the step size of  320 nm over big scanning area. (b) - (d) NEMS photothermal microscopy image of the zoomed-in region indicated by the white box in (a) with 320 nm, 160 nm and 80 nm step sizes, respectively. (e) The corresponding SEM image of the zoomed-in region.}
	\label{fig:fig2} 
\end{figure*}
Gold nanoparticles (AuNPs) with a diameter of \SI{200}{\nano\meter} are first spin-coated and moved into a straight reference line by means of an atomic-force microscope, in order to get a standard sample for system optimizations and calibrations, as shown in Figure~\ref{fig:fig2}. A rough scan with the \SI{633}{\nano\meter} laser over a bigger area in the center is first performed to locate the AuNPs reference line, as shown in Figure~\ref{fig:fig2}a. The AuNPs reference line remains straight through the image, which is an evidence of good stability and negligible drift of the scanning system. We then performed scans within a smaller region centered at the reference line with different scanning step of \SI{320}{\nano\meter}, \SI{160}{\nano\meter}, and \SI{80}{\nano\meter}. The averaged beam radius ($r$) extracted from a Gaussian fit is \SI{800}{\nano\meter}, which is quite close to the nominal beam radius around \SI{750}{\nano\meter} of the objective. (Mitutoyo 50x, 0.55 N.A.) By comparing the NEMS photothermal microscopy images with the reference scanning electron microscopy (SEM) images, single-nanoparticles and aggregates can be identified, as shown in Figure~\ref{fig:fig2}c. With the scanning beam power ($P$) of \SI{85}{\micro\watt}, we can define the responsivity ($\delta R$) as the relative frequency shift $\Delta f/f_0$ per absorbed power by a single AuNP with an absorption cross-section $\sigma_{abs}$ as
\begin{equation}
	\delta R = \frac{\Delta f}{f_0} \frac{\pi r^2}{2P\sigma_{abs}}
\end{equation}
where $f_0$ is the resonance frequency of the NEMS resonator. The  theoretical absorption cross-section of a \SI{200}{\nano\meter} AuNPs based on Mie theory becomes $\sigma_{abs}=$\SI{9e-14}{\meter\squared} \cite{bohren2008absorption,myroshnychenko2008modelling,chien2018single}.  A responsivity of $\delta R=$\SI{7890}{\per\watt} is extracted, which is around two times higher comparing to membranes of the same stress of around \SI{150}{\mega\pascal} and same window size of \SI{1}{\milli\meter} \cite{kurek2017nanomechanical,chien2018single}.

\begin{figure}
	\centering
	\includegraphics[width=0.5\textwidth]{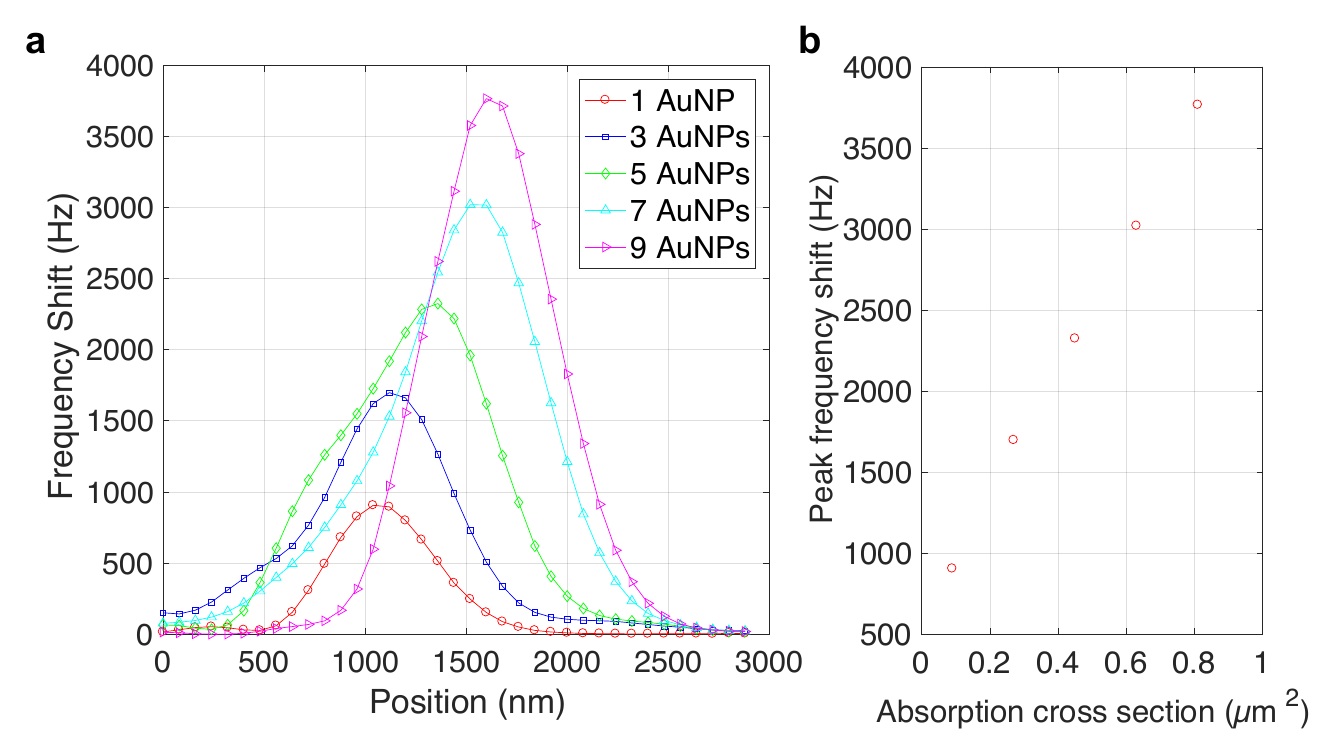} 
	\caption{(a) The frequency shift profiles from different AuNPs aggregates. (b) The peak frequency shift of different aggregates with respect to different absorption cross section.}
	\label{fig:fig4} 
\end{figure} 

The frequency shifts from different AuNPs aggregates extracted from Figure~\ref{fig:fig2}d are plotted in Figure~\ref{fig:fig4}a. The arrangement of the aggregates has slight influences on the scanning profile, which can be seen in the bump overlaying on the Gaussian functions, since the image results from the convolution between beam profile and the absorbing nanoparticle. In general, the peak frequency shift scales linearly with the nubmer of particles per aggregate, as plotted in Figure~\ref{fig:fig4}b. The effect of the plasmonic coupling is not so dominant for the AuNPs at this wavelength. As a result, NEMS photothermal microscopy can effectively identify the number of particles per aggregates, which can be useful for biochemical quantification purposes.

\begin{figure}
	\centering
	\includegraphics[width=0.5\textwidth]{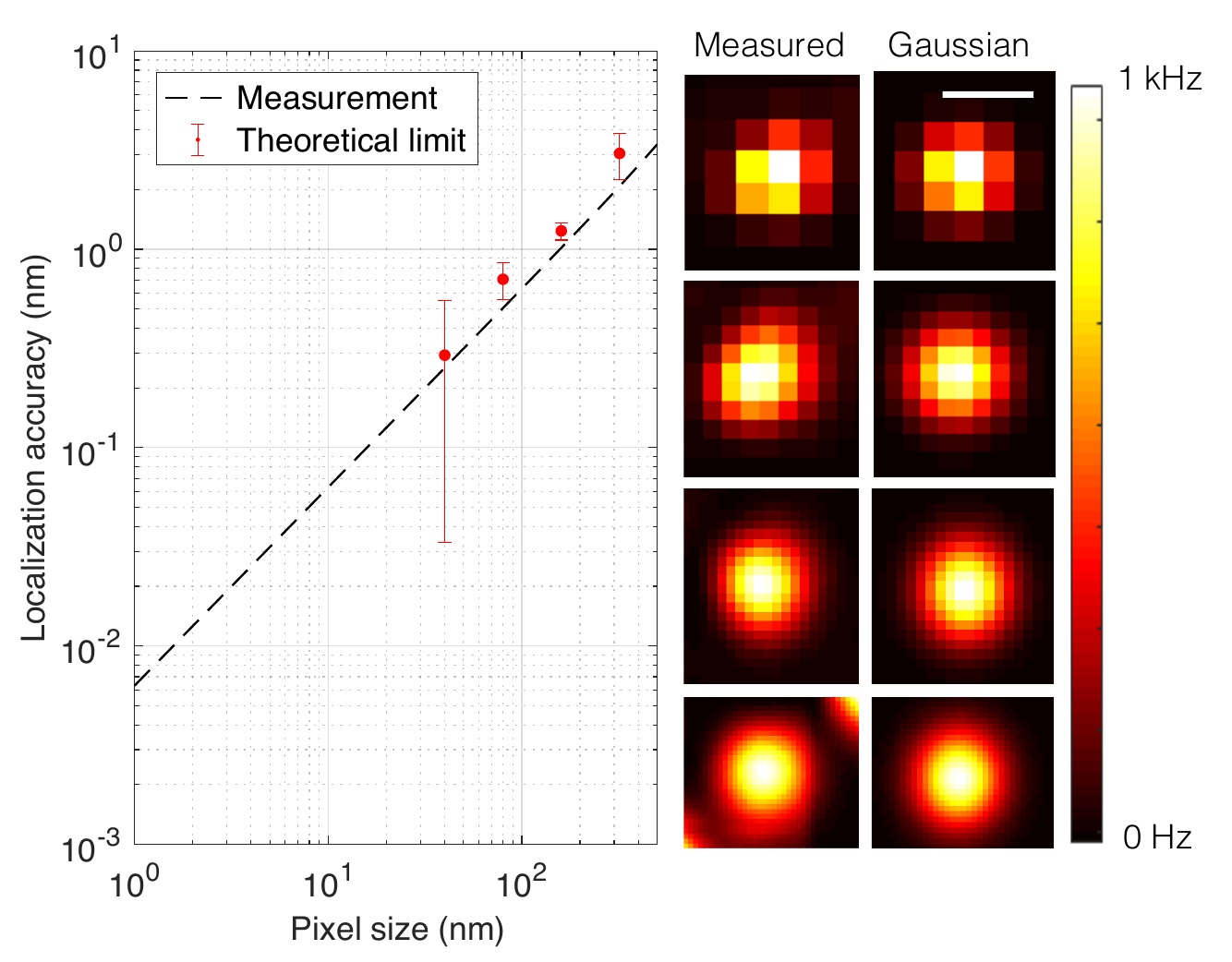} 
	\caption{The localization precision extracted from NEMS photothermal microscopy images with different pixel sizes. The measured and two-dimensional Gaussian-fitted beam profile is plotted on the right columns. The scale bar in the upper-right corner is \SI{1}{\micro\meter}.}
	\label{fig:fig3} 
\end{figure}

From Figure~\ref{fig:fig2}b-d, as the pixel size reduces, the AuNPs reference line can be more clearly resolved, and the center positions of the nanoparticles can be more precisely identified. To discuss the effect of pixel size systematically, the localization precision is calculated for different imaging pixel sizes after fitting with a two-dimensional Gaussian point spread function, as shown in Figure~\ref{fig:fig3}. The localization precision ($\Delta x$) can be expressed as \cite{thompson2002precise,ober2004localization}
\begin{equation}
<(\Delta x)^2> = \frac{s^2 + a^2/12}{N} + \frac{4\sqrt{\pi} s^3 b^2 }{aN^2},
\end{equation}
where $s$ is the standard deviation of the Gaussian function, $a$ is the size of the pixels, and $b$ is the background noise from the images.  $N$ is the sum of the frequency shift levels resulting from the target absorbers, obtained by normalizing the frequency shift with the frequency noise, corresponding to the conventional definition of total photon counts. 

The theoretical limits in Figure~\ref{fig:fig3} are calculated based on perfect focusing with the nominal beam radius of the objective, perfect Gaussian point spread function, and the condition of no background noise of the image ($b$ = 0). In general, the images of AuNPs demonstrate a beam profile which is very close to Gaussian function, as shown in the right columns of Figure~\ref{fig:fig3}. Both the measurements and theoretical limit have shown an improved localization accuracy with smaller pixel sizes. This improvement mainly comes from the increase in the sum of the frequency shift levels ($N$), making the beam profile better-defined with decreased pixel sizes, while the standard deviations of the Gaussian beam profile from different pixel sizes generally remain constant. With a pixel size of \SI{40}{\nano\meter}, a localization accuracy of \SI{3}{\angstrom} can be achieved, almost one order of magnitude better than with a pixel size of \SI{320}{\nano\meter}, which is also the pixel size used in previous work \cite{chien2018single}. This implies the single-molecule localization accuracy of \SI{32}{\nano\meter}, which is lower due to the higher background noise, can be potentially improved by one order of magnitude by using a finer pixel size of \SI{40}{\nano\meter}. The outstanding localization accuracy of the AuNPs in combination with the exceptional chemical stability also make it feasible for diffusion tracking \cite{ando2018single} as well as a reliable component for drift correction and alignment in super-resolution microscopy \cite{bon2015three}.

\begin{figure}
	\centering
	\includegraphics[width=0.5\textwidth]{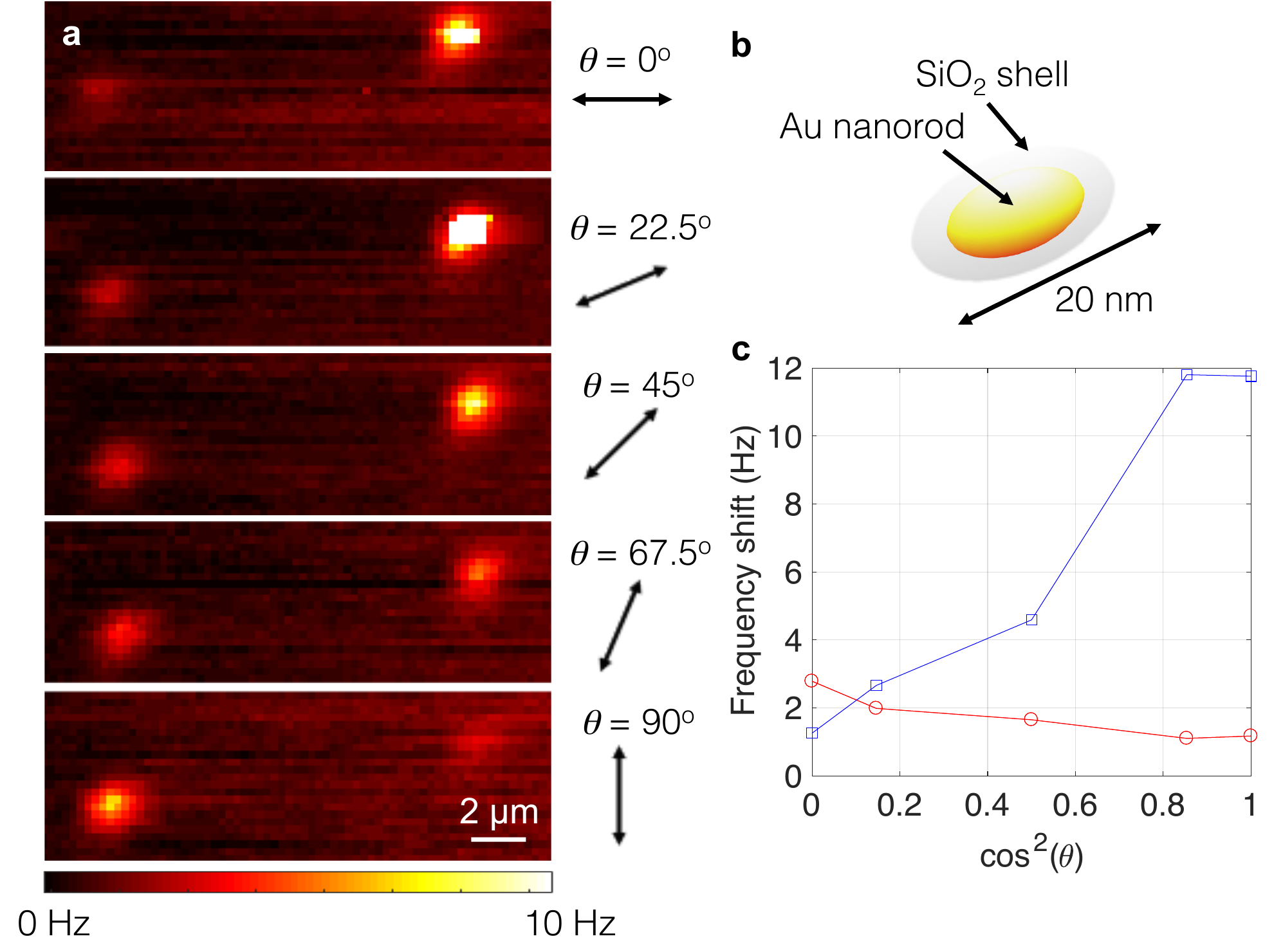} 
	\caption{(a) Nanomechanical photothermal scanning of gold nanorod with polarization angles of 0, 22.5, 45, 67.5 and 90 degree. (b) Schematic of the Au nanorod nanostructure. (c) The peak intensity of the frequency shift for both sets of nanorods in (a).}
	\label{fig:fig5} 
\end{figure} 

To investigate the capability of detecting polarization-dependent absorption, silica-coated gold nanorods with a length of around \SI{20}{\nano\meter} are spin-coated on the NEMS trampoline resonator and scanned with the linearly-polarized Titanium-Sapphire laser locked at \SI{820}{\nano\meter}, as shown in Figure~\ref{fig:fig5}. The absorption peak for the longitudinal polarization of silica-coated gold nanorods is around \SI{820}{\nano\meter}, and around \SI{514}{\nano\meter} for transverse polarization. As a result, maximum absorption and thus frequency shift would be achieved when the nanorods arrange in the same orientation with the beam polarization. Two sets of nanorods with different orientations are located on the trampoline resonator, and the NEMS photothermal microscopy images are obtained with different beam polarizations, as shown in Figure~\ref{fig:fig5}a. The frequency shift of one set of nanorods increases as the polarization angle increases, while the frequency shift of the other set of nanorods decreases. The peak frequency shift of the two nanorods sets are plotted in Figure~\ref{fig:fig5}c with respect to cosine of the polarization angle. Since the amount of nanorods is not the same for both sets, the highest frequency shift is not identical. Interestingly, it can still be clearly identified that the average orientations from this two sets of nanorods are almost opposite. This shows another possibility of NEMS photothermal microscopy for identifying polarization-dependent nano-objects. Furthermore, the polarization modulation imaging has been shown to improve the localization accuracy in fluorescence nanoscopy \cite{hafi2014fluorescence}. As a result, this capability can open another door for the optimization of NEMS photothermal microscopy as a localization microscopy.

\section{Conclusions}
We demonstrate the optimization of NEMS photothermal microscopy with a dedicated optical setup for better beam quality and more control of the beam conditions as a first step towards localization microscopy. The effect of scanning step size on the localization accuracy is discussed systematically, and an optimal localization accuracy of \SI{3}{\angstrom} is achieved for \SI{200}{\nano\meter} AuNPs with low excitation beam power of \SI{85}{\micro\watt} and scanning step of \SI{40}{\nano\meter}. This exceptional localization accuracy along with the chemical stability of the AuNPs enable this system for diffusion tracking \cite{ando2018single} and drift-correction components in super-resolution microscopy simply by spin-coating \cite{bon2015three}. The detection of the polarization-dependent absorption is also demonstrated with silica-coated gold nanorods, which can potentially boost the localization accuracy of NEMS photothermal microscopy. NEMS photothermal microscopy provide a non-fluorescent alternative for nano-object imaging and localization, and would benefit multiple research domains for microscopic analysis.

\begin{acknowledgments}
We gratefully acknowledge the assistance of Sophia Ewert and Patrick Meyer with the sample fabrication and preparation, and the assistance of Florian Patocka with the atomic-force microscopy. This work is supported by the European Research Council under the European Unions Horizon 2020 research and innovation program (Grant Agreement-716087-PLASMECS). 
\end{acknowledgments}

\bibliography{photothermal_NEMS_microscopy.bib}

\end{document}